\documentclass[12pt]{article}
\def\pd{\partial}
\def\mc{\mathcal}

\usepackage[dvips]{graphicx}
\usepackage{amssymb}
\usepackage{amssymb,amsmath}
\usepackage{cite}
\numberwithin{equation}{section}
\numberwithin{table}{section}
\def\AdS{\textrm{AdS}}

\begin{document}
\begin{titlepage}
\begin{center}
\rightline{\small ZMP-HH/16-27}

\vskip 1cm

{\Large\textbf{Supersymmetric $AdS_6$ vacua in six-dimensional
$N=(1,1)$ gauged supergravity}}
\vspace{.8 cm}

{\large\textbf{Parinya Karndumri}}$^a$ and {\large\textbf{Jan
Louis}}$^{b,c}$

\vskip 0.5cm

$^a${\em String Theory and Supergravity Group, Department of Physics,
Faculty of Science, Chulalongkorn University, 254 Phayathai Road,
Pathumwan, Bangkok 10330, Thailand}\\[1ex]
$^b${\em Fachbereich Physik der Universit\"{a}t Hamburg, Luruper Chaussee 149, 22761 Hamburg, Germany}\\[1ex]
$^c${\em Zentrum f\"{u}r Mathematische Physik, Universit\"{a}t Hamburg,
Bundesstrasse 55, D-20146 Hamburg, Germany}

\vskip 0.2cm

email: {\tt parinya.ka@hotmail.com, jan.louis@desy.de }

\vspace{1.5 cm}
\end{center}
We study fully supersymmetric $AdS_6$ vacua of
half-maxi\-mal $N=(1,1)$  gauged supergravity in six space-time dimensions coupled to $n$
vector multiplets. We show that the existence of $AdS_6$ backgrounds requires
that the gauge group is of the form $G'\times
G''\subset SO(4,n)$ where $G'\subset SO(3,m)$ and $G''\subset
SO(1,n-m)$. In the $AdS_6$ vacua this gauge group is broken to its maximal
compact subgroup $SO(3)\times H'\times H''$ where
$H'\subset SO(m)$ and $H''\subset SO(n-m)$.
Furthermore, the $SO(3)$ factor is the R-symmetry gauged by three of the four
graviphotons.  We further show that the
$AdS_6$ vacua have no moduli that preserve all supercharges. This is
precisely in agreement with the absence of supersymmetric marginal deformations in
holographically dual five-dimensional superconformal field theories.
\vfill
December 2016
\end{titlepage}

\section{Introduction}
Supersymmetric anti-de Sitter (AdS) vacua and their moduli spaces
of gauged supergravities are of particular interest in
the AdS/CFT correspondence \cite{maldacena}. The
AdS vacua correspond to conformal fixed points of the holographically dual field
theories while the moduli spaces describe the conformal manifolds near
these fixed points \cite{Aharony:2002hx,Kol:2002zt}. The latter encode useful information about the exactly
marginal deformations of the corresponding superconformal field
theories (SCFTs).

AdS backgrounds of gauged supergravities and their moduli spaces have been studied
in various space-time dimensions  with different numbers of supercharges.
In this paper we exclusively focus on
 the half-maximal gauged $N=(1,1)$ supergravity in six space-time dimensions ($d=6$)
 and their maximally supersymmetric $\AdS_6$ backgrounds.\footnote{In fact,
 the $N=(1,1)$ supergravity
 is  the only gauged supergravity in $d=6$
that admits maximally supersymmetric $\AdS_6$ backgrounds \cite{Max_super_bg_Jan}.
This in turn is consistent with the
known classification of the AdS superalgebras given in
\cite{Nahm_AdS_algebra}.}
 This supergravity is also known as $F(4)$ supergravity and was first constructed in
  \cite{F4_Romans}.  It is
non-chiral and can be coupled to an
arbitrary number $n$ of vector multiplets.
 Each vector multiplet
contains four scalars and
together with the dilaton in the gravity
multiplet, they  parametrize  the $(4n+1)$-dimensional
coset manifold
$\mathbb{R}^+\times SO(4,n)/SO(4)\times SO(n)$.
The corresponding gauged supergravity was
constructed in \cite{F4SUGRA1,F4SUGRA2} by extending the pure $F(4)$
supergravity using the geometric group manifold approach.
 \cite{F4SUGRA1,F4SUGRA2} also showed that for a gauge group $SU(2)_R\times G$
and $G\subset SO(n)$ a maximally supersymmetric $\AdS_6$
vacuum exists where the full  $SU(2)_R\times G$ symmetry is realized at the origin of the scalar
manifold. This vacuum could be identified with the near horizon
geometry of the D4-D8 brane system \cite{D4D8}. For the case of
$n=3$ vector multiplets and $G=SO(3)$, another non-trivial $\AdS_6$
vacuum breaking the $SU(2)_R\times SO(3)$ symmetry to
$SO(3)_{\textrm{diag}}$ and preserving the full $N=(1,1)$
supersymmetry has been identified in \cite{F4_flow}.

 In this paper we do not specify the gauge group upfront but instead follow the strategy
 of \cite{AdS4_moduli,AdS4_N4_Jan,AdS_7_N2_Jan,AdS5_N4_Jan,AdS5_N2_Jan}
in that we first determine the general conditions on the parameters of the gauged supergravity  such that $\AdS_6$ backgrounds which preserve all supercharges can exist. In half-maximal supergravities it is then possible to also give all possible gauge groups that can have such vacua.
Concretely we find that the gauge group has to be of the form $G'\times
G''\subset SO(4,n)$ where $G'\subset SO(3,m)$ and $G''\subset
SO(1,n-m)$. In the $AdS_6$ vacua this gauge group is broken to its maximal
compact subgroup $SO(3)\times H'\times H''$ where
$H'\subset SO(m)$ and $H''\subset SO(n-m)$.
The $SO(3)\sim SU(2)$ factor precisely is the R-symmetry and it is gauged by three of the four
graviphotons.
Finally, we derive the necessary conditions for the existence of a supersymmetric moduli space near these vacua. For the case at hand we find that no moduli space is possible
which is again consistent with the fact that the holographically dual SCFTs have no
supersymmetric exactly marginal deformations.

In the AdS/CFT correspondence, $\AdS_6$ vacua are also
relevant for studying strongly coupled five-dimensional SCFTs
arising from the dynamics of D4-D8 branes
\cite{Seiberg_5Dfield,D4D8}.
The interpretation in terms of
$\AdS_6$ geometry in \cite{ferrara_AdS6} inspired various studies
considering gravity duals of these SCFTs including a recent
generalization to quiver gauge theories in \cite{Bergman}. Finding
$\AdS_6$ solutions in type II and eleven dimensional supergravities
also deserves detailed investigations.\footnote{See
\cite{AdS6_from10D,AdS6_Tomasiello,AdS_6_Kim} for recent results
along this direction and references therein.} In this paper, however, we stay in $d=6$ throughout
the analysis leaving the higher dimensional origins of these
vacua for future work.

 The paper is organized as follow. In section \ref{F4_SUGRA}, we set the stage for our analysis
and recall the relevant features of $N=(1,1)$ gauged supergravity.
The conditions for the existence of maximally
supersymmetric $\AdS_6$ vacua are then derived in section
\ref{AdS6_vacua}, and the analysis of the moduli space is carried
out in section \ref{moduli}. We finally end the paper by giving some
conclusions and comments on the results in section \ref{conclusion}.

\section{$N=(1,1)$ gauged supergravity in six dimensions}\label{F4_SUGRA}
In this section, we briefly review $N=(1,1)$ gauged supergravity
coupled to $n$ vector multiplets in order to set up the notation for the
later analysis. More details on this gauged
supergravity can be found in \cite{F4SUGRA1,F4SUGRA2}. We
will follow most of the conventions in these two references.

 The possible supermultiplets are the gravitational multiplet  and $n$ vector multiplets given
respectively by
\begin{equation}
\left(e^a_\mu,\psi^A_\mu, A^\alpha_\mu, B_{\mu\nu}, \chi^A,
\sigma\right)\qquad \textrm{and}\qquad
(A_\mu,\lambda_A,\phi^\alpha)^I\, .
\end{equation}
The bosonic fields of the supergravity multiplet are given by the
graviton $e^a_\mu$, the dilaton $\sigma$, four
graviphotons $A^\alpha_\mu$, and a two-form field $B_{\mu\nu}$ while each vector multiplet contains a vector, $A_\mu$, and four scalars, $\phi^\alpha$. Two
sets of indices $\alpha,\beta,\ldots=0,1,2,3$ and $I,J,\ldots=1,\ldots ,n$ label
the $n+4$ vector fields. Space-time and tangent space indices are
denoted respectively by $\mu,\nu=0,\ldots ,5$ and $a,b=0,\ldots, 5$.
We will also follow the mostly minus space-time signature $(+-----)$
of \cite{F4SUGRA1,F4SUGRA2}.

 The fermionic fields consist of two gravitini $\psi^A_\mu$, two spin-$\frac{1}{2}$ fields $\chi^A$ and $2n$ gauginos $\lambda_A^I$. All of these fields and the supersymmetry parameter $\epsilon^A$ are
eight-component pseudo-Majorana spinors and transform in the fundamental representation of the $SU(2)_R\sim
USp(2)_R$ R-symmetry denoted by indices $A,B=1,2$.

The dilaton and the $4n$ scalars $\phi^{\alpha I}$ of the vector multiplets  span the coset manifold
\begin{equation}
\mathbb{R}^+\times SO(4,n)/SO(4)\times SO(n)\, .
\end{equation}
The second factor can in turn be parametrized by the coset
representative $L^\Lambda_{\phantom{\Lambda}\Sigma}$ with
$\Lambda,\Sigma,\ldots=1,2,\ldots, n+4$. It is convenient to split the
indices transforming under the compact group $SO(4)\times SO(n)$ as
$\Lambda=(\alpha,I)$ and further under the $SO(3)_R\times SO(n)$ as
$\Lambda=(0,r,I)$ with $r,s,\ldots=1,2,3$. The $SO(3)_R$ is
identified with the diagonal subgroup of $SO(3)\times SO(3)\sim
SO(4)$. The coset representative can be accordingly decomposed as
\begin{equation}
L^\Lambda_{\phantom{\Lambda}\Sigma}=(L^\Lambda_{\phantom{\Lambda}\alpha},
L^\Lambda_{\phantom{\Lambda}I})=(L^\Lambda_{\phantom{\Lambda}0},
L^\Lambda_{\phantom{\Lambda}r},L^\Lambda_{\phantom{\Lambda}I})\, .
\end{equation}
Furthermore, all of the $n+4$ vector fields will be collectively
denoted by $A^\Lambda_\mu=(A^0_\mu,A^r_\mu,A^I_\mu)$. Being
$SO(4,n)$ matrices, the $L^\Lambda_{\phantom{\Lambda}\Sigma}$ satisfy
the relation
\begin{equation}
\eta_{\Lambda\Sigma}=L^0_{\phantom{\Lambda}\Lambda}L^0_{\phantom{\Lambda}\Sigma}
+L^i_{\phantom{\Lambda}\Lambda}L^i_{\phantom{\Lambda}\Sigma}-
L^I_{\phantom{\Lambda}\Lambda}L^I_{\phantom{\Lambda}\Sigma}\label{SO4_n_identity}
\end{equation}
with $\eta_{\Lambda\Sigma}=(1,1,1,1,-1,-1,\ldots, -1)$.

 We now turn to the gauged version of this 
 supergravity.
The most complete gauged $N=(1,1)$ supergravity up to now is given
in \cite{F4SUGRA1,F4SUGRA2}. As in seven dimensions, the full
$SO(4,n)$ covariant formulation in terms of the embedding tensors has
not been worked out yet although the corresponding components of the
embedding tensor have been identified in \cite{Eric_Kac_Moody} using
the Kac-Moody approach. In this paper, we will restrict ourselves to
the gauged supergravity constructed in \cite{F4SUGRA1,F4SUGRA2}.

 Gauging is implemented by making a particular subgroup $G$
of $SO(4,n)$ local such that the adjoint representation of $G$ can
be embedded in the fundamental representation, $\mathbf{n+4}$, of
$SO(4,n)$, and $\eta_{\Lambda\Sigma}$ contains the Cartan-Killing
form of the gauge group. Consistency with supersymmetry requires
that the 
structure constants are totally anti-symmetric, i.e.\
$f_{\Lambda\Sigma\Pi}=f_{\Lambda\Sigma}^{\phantom{\Lambda\Sigma}\Gamma}\eta_{\Gamma\Pi}=f_{[\Lambda\Sigma\Pi]}$.
In the embedding tensor formalism, this condition is called the
linear constraint.

The $f_{\Lambda\Sigma}^{\phantom{\Lambda\Sigma}\Gamma}$ appear as
structure constants in the gauge algebra
\begin{equation}
\left[T_\Lambda,T_\Sigma\right]=f_{\Lambda\Sigma}^{\phantom{\Lambda\Sigma}\Gamma}T_\Gamma
\end{equation}
in which $T_\Lambda$ are gauge generators. These structure constants
must satisfy the Jacobi identity
\begin{equation}
f_{[\Sigma\Gamma}^{\phantom{\Sigma\Gamma}\Delta}f_{\Lambda]\Delta}^{\phantom{\Lambda\Delta}\Pi}=0\label{Jacobi}
\end{equation}
which in the embedding tensor formalism is the so-called quadratic
constraint. In general, this constraint comes from the requirement
that the gauge generators, obtained from appropriate projections of
the global symmetry generators by the embedding tensor, form a
closed Lie algebra of the corresponding gauge group.

The bosonic Lagrangian with only the metric and
scalars non-vanishing reads
\begin{equation}
e^{-1}\mc{L}=-\frac{1}{4}R+\pd_\mu \sigma \pd^\mu \sigma+\frac{1}{4}(P^{I0}_\mu P^\mu_{I0}+P^{Ir}_\mu P^{\mu}_{Ir})-V\ ,
\end{equation}
where the scalar kinetic term is written in terms of the Maurer-Cartan one-forms
\begin{equation}
P^I_{\phantom{I}0}=(L^{-1})^I_{\phantom{I}\Lambda}(dL^\Lambda_{\phantom{\Lambda}0}+f^\Lambda_{\phantom{\Lambda}\Gamma\Pi}A^\Gamma L^\Pi_{\phantom{\Lambda}0}),\qquad P^I_{\phantom{I}r}=(L^{-1})^I_{\phantom{I}\Lambda}(dL^\Lambda_{\phantom{\Lambda}r}+f^\Lambda_{\phantom{\Lambda}\Gamma\Pi}A^\Gamma L^\Pi_{\phantom{\Lambda}r})\, .
\end{equation}
The scalar potential $V$ is given by
\begin{equation}\begin{aligned}\label{Vdef}
V&=-5\left[\tfrac{1}{144}(Ae^\sigma+6me^{-3\sigma}L_{00})^2+\tfrac{1}{16}(B_ie^\sigma-2me^{-3\sigma}L_{0i})^2\right]\\
&+\tfrac{1}{144}(Ae^\sigma-18me^{-3\sigma}L_{00})^2+\tfrac{1}{16}(B_ie^\sigma+6me^{-3\sigma}L_{0i})^2\\
&+\tfrac{1}{4}(C^I_{\phantom{I}t}C_{It}+4D^I_{\phantom{I}t}D_{It})e^{2\sigma}-m^2e^{-6\sigma}L_{0I}L^{0I}
\end{aligned}\end{equation}
where $m$ is the mass of the two-form in the gravitational multiplet and we abbreviated
\begin{equation}\begin{aligned}\label{ABCD}
A&=\epsilon^{rst}K_{rst}\ ,\qquad B^i=\epsilon^{ijk}K_{jk0}\ ,\\
C_I^{\phantom{I}t}&=\epsilon^{trs}K_{rIs}\ ,\qquad D_{It}=K_{0It}\ ,
\end{aligned}
\end{equation}
with the ``dressed'' structure constants given by
\begin{equation}\begin{aligned}\label{Kabbre}
K_{rst}&=f_{\Lambda \Sigma\Pi}L^\Lambda_{\phantom{r}r}(L^{-1})_s^{\phantom{s}\Sigma}L^\Pi_{\phantom{s}t},\qquad K_{rs0}=f_{\Lambda \Sigma\Pi}L^\Lambda_{\phantom{r}r}(L^{-1})_s^{\phantom{s}\Sigma}L^\Pi_{\phantom{s}0}, \\
K_{rIt}&=f_{\Lambda \Sigma\Pi}L^\Lambda_{\phantom{r}r}(L^{-1})_I^{\phantom{s}\Sigma}L^\Pi_{\phantom{s}t},\qquad K_{0It}=f_{\Lambda \Sigma\Pi}L^\Lambda_{\phantom{r}0}(L^{-1})_I^{\phantom{s}\Sigma}L^\Pi_{\phantom{s}t} \, .
\end{aligned}
\end{equation}

 The supersymmetry transformations of the fermions which will play an
important role in the following analysis are given by
\begin{equation}\begin{aligned}\label{fermiontrans}
\delta \psi_{A\mu}&=D_\mu \epsilon_A+S_{AB}\gamma_\mu \epsilon^B,\\
\delta \chi_A &=\tfrac{i}{2}\gamma^\mu\pd_\mu \sigma \epsilon_A+N_{AB}\epsilon^B,\\
\delta\lambda^I_A&=-iP^I_{ri}\sigma^{rAB}\pd_\mu \phi^i \gamma^\mu \epsilon_B+iP^I_{0i}\epsilon^{AB}\pd_\mu \phi^i\gamma^7\gamma^\mu \epsilon_B+M^I_{AB}\epsilon^B,
\end{aligned}
\end{equation}
where the fermion-shift matrices are defined as
\begin{equation}\begin{aligned}\label{shiftmatrices}
S_{AB}&=\tfrac{i}{24}\left[Ae^\sigma+6me^{-3\sigma}(L^{-1})_{00}\right]\epsilon_{AB}
-\tfrac{i}{8}\left[B_te^\sigma-2me^{-3\sigma}(L^{-1})_{t0}\right]\gamma^7\sigma^t_{AB},\\[1ex]
N_{AB}&=\tfrac{1}{24}\left[Ae^\sigma-18me^{-3\sigma}(L^{-1})_{00}\right]\epsilon_{AB}
+\tfrac{1}{8}\left[B_te^\sigma+6me^{-3\sigma}(L^{-1})_{t0}\right]\gamma^7\sigma^t_{AB},\\[1ex]
M^I_{AB}&=(-C^I_{\phantom{I}t}+2i\gamma^7D^I_{\phantom{I}t})e^\sigma\sigma^t_{AB}-2me^{-3\sigma}(L^{-1})^I_{\phantom{I}0}\gamma^7\epsilon_{AB}\, .
\end{aligned}
\end{equation}
In the present convention, the anti-symmetric matrix
$\epsilon_{AB}=-\epsilon_{BA}$ is taken to be $\epsilon_{12}=\epsilon^{12}=1$. The
$\sigma^t_{AB}$ matrices are related to the usual Pauli matrices
$\sigma^{tA}_{\phantom{tA}B}$ by the relation\footnote{Note that $\sigma^t_{AB}=\sigma^t_{(AB)}$.}
\begin{equation}
\sigma^t_{AB}=\sigma^{tC}_{\phantom{tA}B}\epsilon_{CA}\, .
\end{equation}
Finally,
the chirality matrix $\gamma_7$ is defined by
\begin{equation}
\gamma_7=i\gamma_0\gamma_1\gamma_2\gamma_3\gamma_4\gamma_5
\end{equation}
with $\gamma^2_7=-\mathbb{I}$ and $\gamma_7^T=-\gamma_7$.

\section{Maximally supersymmetric $\AdS_6$ vacua}\label{AdS6_vacua}
We now determine the maximally supersymmetric $\AdS_6$ vacua preserving all sixteen supercharges. In order to do so, we impose that the following conditions vanish for all supercharges in the background
\begin{equation}\label{dPsi}
\langle \delta \psi_{\mu A}\rangle=0\ ,\qquad
\langle \delta \chi_A\rangle=0\ ,\qquad
\langle \delta\lambda^I_A\rangle=0\ .
\end{equation}
Due to the symmetries of $\sigma^t_{AB}=\sigma^t_{(AB)}$ and
$\epsilon_{AB}=\epsilon_{[AB]}$, the linear independence of
$\gamma_7$ and $\mathbb{I}$ and by using \eqref{fermiontrans} and \eqref{shiftmatrices}
we infer that the second and third equations imply
\begin{eqnarray}
\langle Ae^\sigma-18me^{-3\sigma}(L^{-1})_{00}\rangle &=&0,\label{A_condition} \\
\langle e^{-3\sigma}(L^{-1})^I_{\phantom{I}0}\rangle &=&0,\label{L0I_condition}\\
\langle B_te^\sigma+6me^{-3\sigma}(L^{-1})_{t0}\rangle&=&0, \label{B_condition1}\\
\langle C^I_{\phantom{I}t}\rangle=\langle
D^I_{\phantom{I}t}\rangle&=&0\, .\label{AdS_conditions1}
\end{eqnarray}
From \eqref{ABCD} we learn that the first condition in
\eqref{AdS_conditions1} is equivalent to
\begin{equation}
\epsilon^{trs}K_{rIs}=0\, .
\end{equation}
The second condition in
\eqref{AdS_conditions1} yields $K_{0It}=0$ so that together we have
\begin{equation}
K_{rIs}=K_{0It}=0\, .\label{D_C_conditions}
\end{equation}
Using \eqref{ABCD}  we can rewrite condition \eqref{A_condition}  as
\begin{equation}
\epsilon^{rst}K_{rst}=18me^{-4\langle \sigma\rangle}\langle(L^{-1})_{00}\rangle\label{eqx}
\end{equation}
which is solved by
\begin{equation}\label{Ksol}
K_{rst}=g\epsilon_{rst}
\end{equation}
for an arbitrary $SU(2)_R$ gauge coupling $g$.
We can accordingly
determine the background value of the dilaton by inserting \eqref{Ksol}  into \eqref{eqx}
\begin{equation}
e^{-4\langle \sigma\rangle}\langle
(L^{-1})_{00}\rangle=\frac{g}{3m}\, .
\end{equation}
The remaining conditions \eqref{L0I_condition} and \eqref{B_condition1} give
\begin{equation}
\langle (L^{-1})^I_{\phantom{I}0}\rangle=0,\qquad \langle B_t\rangle=-6me^{-4\langle\sigma\rangle}\langle(L^{-1})_{t0}\rangle\, .\label{AdS_condition1}
\end{equation}
 Using the component-$(0I)$ and -$(0i)$ of the relation
\eqref{SO4_n_identity} and the identity $L^{-1}=\eta L^T\eta$, we
find that $L_{0I}=0$ implies $L_{0i}=0$ and thus
\begin{equation}
\langle B_t\rangle=0 \ .
\end{equation}
Using the definition of $B_t$ given in \eqref{ABCD} we thus arrive at
\begin{equation}
K_{jk0}=0\, .\label{B_condition}
\end{equation}
 By taking the $(00)$-component of the relation
\eqref{SO4_n_identity}, we find that  $L_{0I}=L_{0i}=0$ also
implies $L_{00}=1$. Inserting the results obtained so far into \eqref{Vdef}
we conclude that the background value of the scalar potential (related to the cosmological
constant) in an $\AdS_6$ vacuum is given by
\begin{equation}
\langle V\rangle =-20m^2\left(\frac{g}{3m}\right)^\frac{3}{2}\, .
\end{equation}
We see that $\AdS$ vacua do not exist for $m=0$ as
already pointed out in \cite{F4SUGRA1,F4SUGRA2}.\footnote{Recall that $m$ is the mass parameter of the two-form in the gravitational multiplet.} This is very
similar to AdS backgrounds of half-maximal supergravities in seven dimensions
\cite{AdS_7_N2_Jan,7D_flow,Non_compact_7DN2}. Note also that by
shifting the value of $\langle\sigma\rangle$ we can choose $g=3m$ as
in \cite{F4SUGRA1,F4SUGRA2}.

In order to continue let us recall that we are left with the unconstrained structure constants
\begin{equation}\label{Knt}
K_{rst}\ ,\qquad K_{rIJ}\ ,\qquad K_{0IJ}\ , \qquad K_{IJK}\ ,
\end{equation}
whose choice specify the particular supergravity at hand.
We can now use the quadratic constraint to determine the corresponding gauge groups.
These are the gauge groups which can occur in the supergravities that admit maximally supersymmetric $\AdS_6$ vacua.
For the case at hand the quadratic constraint
reduces to the usual Jacobi identity given in \eqref{Jacobi}.

As a warm up let us first consider the simple situation where $K_{rIJ}=K_{0IJ}=K_{IJK}=0$ and we only have $K_{rst}$ non-zero. In this
case, equation \eqref{Jacobi} reduces to the Jacobi identity of
an $SO(3)$ algebra with the structure constants
$K_{rst}=g\epsilon_{rst}$. We then simply recover the pure $F(4)$
gauged supergravity with an $SU(2)\sim SO(3)$ gauge group constructed
in \cite{F4_Romans}.

For $K_{rIJ}=K_{0IJ}=0$ but $K_{IJK}\neq 0$, the condition
\eqref{Jacobi} gives rise to two separate Jacobi identities for
$K_{rst}$ and $K_{IJK}$ which correspond to two commuting compact groups. The
gauge group is accordingly $SO(3)\times H$ with
$H\subset SO(n)$ and compact. This gauge group and the resulting $\AdS_6$ vacuum
together with the dual five-dimensional SCFT have already been
studied in \cite{F4SUGRA1,F4SUGRA2}.

As a next step let us also take $K_{rIJ}\neq 0$ but still have $K_{0IJ}=0$. In this case
the $SO(3)$-singlet graviphoton $A^0$ decouples from all other gauge bosons. This
is very similar to the seven-dimensional case studied in
\cite{AdS_7_N2_Jan} where the gauge groups are embedded in
$SO(3,n)\subset SO(4,n)$.
If one additionally assumes that the gauge group is semi-simple one can in fact list all possibilities.
The Cartan-Killing form of these gauge groups must be embeddable in the
$SO(3,n)$ invariant tensor $\eta=(\delta_{rs},-\delta_{IJ})$ which
imposes a strong constraint.
Furthermore, the existence of supersymmetric $\AdS_6$ vacua requires
that the gauge groups must contain $SO(3)$ as a subgroup. The only
possible semisimple gauge groups are then given by
\begin{equation}
\tilde{G}\times H
\end{equation}
where $\tilde{G}=SO(3),SO(3,1)$ or $SL(3,\mathbb{R})$ and $H\subset
SO(n+3-\textrm{dim}\,\tilde{G})$ is a compact group.

We finally consider the most general case with all structure constants in \eqref{Knt} non-zero.
Follow a similar analysis in \cite{AdS5_N4_Jan} we
introduce the gauge generators embedded in $SO(4,n)$ as
\begin{equation}
(T_\Lambda)_{\Gamma}^{\phantom{\Gamma}\Pi}
=f_{\Lambda}^{\phantom{\Lambda}\Sigma\Delta}(t_{\Sigma\Delta})_{\Gamma}^{\phantom{\Gamma}\Pi}
=f_{\Lambda\Gamma}^{\phantom{\Lambda\Gamma}\Pi}
\end{equation}
where
$(t_{\Sigma\Delta})_{\Gamma}^{\phantom{\Gamma}\Pi}=\delta^\Pi_{[\Sigma}\eta_{\Delta]\Gamma}$
are $SO(4,n)$ generators in the vector representation.
Splitting the indies $\Lambda=(0,i,I)$ decomposes the gauge
generators as
\begin{equation}
(T_0)_{\Gamma}^{\phantom{\Gamma}\Pi}=f_{0\Gamma}^{\phantom{\Lambda\Gamma}\Pi},\qquad
(T_i)_{\Gamma}^{\phantom{\Gamma}\Pi}=f_{i\Gamma}^{\phantom{\Lambda\Gamma}\Pi},\qquad
(T_I)_{\Gamma}^{\phantom{\Gamma}\Pi}=f_{I\Gamma}^{\phantom{\Lambda\Gamma}\Pi},\qquad
\end{equation}
which couple to the vector fields $A^0$, $A^i$ and $A^I$,
respectively.

It is more convenient to write down the various independent
components of the Jacobi identity. They read
\begin{eqnarray}
K_{[ij}^{\phantom{IJ}l}K_{k]l}^{\phantom{IJ}m}&=&0\ ,\label{Jeq11}\\
K_{iJ}^{\phantom{0I}I}K_{IK}^{\phantom{KJ}j}+K_{Ki}^{\phantom{Ki}I}K_{IJ}^{\phantom{IJ}j}
+K_{JK}^{\phantom{JK}r}K_{ri}^{\phantom{ri}j}&=&0\ ,\label{Jeq12}\\
K_{iJ}^{\phantom{0I}I}K_{IK}^{\phantom{KJ}L}+K_{Ki}^{\phantom{Ki}I}K_{IJ}^{\phantom{IJ}L}+
K_{JK}^{\phantom{JK}I}K_{Ii}^{\phantom{Ii}L}&=&0\ ,\label{Jeq2}\\
K_{0I}^{\phantom{0I}J}K_{Jj}^{\phantom{Kj}K}
+K_{Ij}^{\phantom{0I}J}K_{J0}^{\phantom{Jj}K}&=&0\ ,\label{Jeq31}\\
K_{IJ}^{\phantom{IJ}K}K_{K0}^{\phantom{KJ}L}+K_{0I}^{\phantom{0I}K}K_{KJ}^{\phantom{KJ}L}
+K_{J0}^{\phantom{0I}K}K_{KI}^{\phantom{KI}L}&=&0\ ,\label{Jeq32}\\
K_{[IJ}^{\phantom{IJ}0}K_{K]0}^{\phantom{KJ}M}+K_{[IJ}^{\phantom{IJ}r}K_{K]r}^{\phantom{KJ}M}
+K_{[IJ}^{\phantom{IJ}L}K_{K]L}^{\phantom{KJ}M}&=&0\ .\label{Jeq4}
\end{eqnarray}
The first two relations \eqref{Jeq11}, \eqref{Jeq12} imply that the $SO(3)$ generators
$T_i$ have non-vanishing elements in both $SO(3)$ and $SO(n)$
blocks. We therefore split the indices $I,J,K,\ldots$ into two sets
$\hat{I},\hat{J}, \hat{K}=1,\ldots ,m$ and
$\tilde{I},\tilde{J},\tilde{K}=1,\ldots n-m$ such that only the
$\hat{I},\hat{J}, \hat{K}$ indices mix with $r,s,t$ indices. Or, in
other word, we have $K_{r\hat{I}\hat{J}}\neq 0$ and
$K_{r\tilde{I}\tilde{J}}=0$. With this convention the $SO(3)$ generators take the form
\begin{equation}
T_i=\left(
      \begin{array}{cccc}
        0 &  &  &  \\
         & K_{ij}^{\phantom{ij}k} &  &  \\
         &  & K_{i\hat{J}}^{\phantom{i\hat{J}}\hat{K}} &  \\
         &  &  & 0_{n-m} \\
      \end{array}
    \right)\ ,
\end{equation}
where $0_n$ indicates an $n\times n$ zero matrix.

The relation \eqref{Jeq31} corresponds to $\left[T_i,T_0\right]=0$ and thus
$T_0$ and $T_i$ cannot have
common $I,J,K$ indices or equivalently
$K_{0\hat{I}\hat{J}}=K_{0\tilde{I}\hat{J}}=0$. This determines
the $T_0$ generator to be
\begin{equation}
T_0=\left(
      \begin{array}{cccc}
        0 &  &  &  \\
         & 0_3 &  &  \\
         &  & 0_{m} &  \\
         &  &  & K_{0\tilde{J}}^{\phantom{0\tilde{J}}\tilde{K}} \\
      \end{array}
    \right).
\end{equation}
Equation \eqref{Jeq2} and the $(\hat{I},\hat{J},\hat{K},\hat{M})$
components of relation \eqref{Jeq4} imply that the $T_{\hat{I}}$
generators are given by
\begin{equation}
T_{\hat{I}}=\left(
      \begin{array}{cccc}
        0 &  &  &  \\
         & 0_3 & K_{\hat{I}r}^{\phantom{\hat{I}r}\hat{K}} &  \\
         & K_{\hat{I}\hat{J}}^{\phantom{\hat{I}\hat{J}}r} & K_{\hat{I}\hat{J}}^{\phantom{\hat{I}\hat{J}}\hat{K}} &  \\
         &  &  & 0_{n-m} \\
      \end{array}
    \right).
\end{equation}
Therefore, the $(T_i,T_{\hat{I}})$ generators together form a
non-compact group $G'\subset SO(3,m)$, $m\leq n$.

Finally, the relation \eqref{Jeq32} and the
$(\tilde{I},\tilde{J},\tilde{K},\tilde{M})$ components of relation
\eqref{Jeq4} determine the structure of $T_{\tilde{I}}$ to be
\begin{equation}
T_{\tilde{I}}=\left(
      \begin{array}{cccc}
        0 &  &  & K_{\tilde{I}0}^{\phantom{\tilde{I}0}\tilde{K}} \\
         & 0_3 &  &  \\
         &  & 0_m &  \\
         K_{\tilde{I}\tilde{J}}^{\phantom{\tilde{I}\tilde{J}}0}&  &  & K_{\tilde{I}\tilde{J}}^{\phantom{\tilde{I}\tilde{J}}\tilde{K}} \\
      \end{array}
    \right).
\end{equation}
These generators together with $T_0$ form another non-compact group
$G''\subset SO(1,n-m)$. We then conclude that the general 
gauge group admitting maximally supersymmetric $\AdS_6$ vacua take
the form
\begin{equation}
G'\times G''
\end{equation}
where $G'\subset SO(3,m)$ and $G''\subset SO(1,n-m)$. In an $\AdS_6$
background, the gauge group is broken to its maximal compact
subgroup $SO(3)\times  H'\times H''$ in which $H'\subset SO(m)$ and
$H''\subset SO(n-m)$.

To confirm this, we inspect the massive vector fields arising from
the above symmetry breaking. Defining
$A^{\hat{I}}=(L^{-1})^{\hat{I}}_{\phantom{I}\Lambda}A^\Lambda$ and
$A^{\tilde{I}}=(L^{-1})^{\tilde{I}}_{\phantom{I}\Lambda}A^\Lambda$,
we find that various components of the Maurer-Cartan one-form
$P^I_{\phantom{I}\alpha}$ are given by
\begin{eqnarray}
P^{\hat{I}}_{\phantom{I}0}&=&(L^{-1})^{\hat{I}}_{\phantom{I}\Lambda}dL^\Lambda_{\phantom{\Lambda}0},\qquad
P^{\hat{I}}_{\phantom{I}r}=(L^{-1})^{\hat{I}}_{\phantom{I}\Lambda}dL^\Lambda_{\phantom{\Lambda}r}
+K^{\hat{I}}_{\phantom{I}\hat{J}r}A^{\hat{J}},\nonumber \\
P^{\tilde{I}}_{\phantom{I}0}&=&(L^{-1})^{\tilde{I}}_{\phantom{I}\Lambda}dL^\Lambda_{\phantom{\Lambda}0}
+K^{\tilde{I}}_{\phantom{I}\tilde{J}0}A^{\tilde{J}},\qquad
P^{\tilde{I}}_{\phantom{I}r}=(L^{-1})^{\tilde{I}}_{\phantom{I}\Lambda}dL^\Lambda_{\phantom{\Lambda}r}
\, .
\end{eqnarray}
By computing the scalar kinetic terms, we can indeed see that there
is precisely one massive vector field corresponding to each
non-compact generators
$K_{\hat{I}\hat{J}}^{\phantom{\hat{I}\hat{J}}r}$ and
$K_{\tilde{I}\tilde{J}}^{\phantom{\tilde{I}\tilde{J}}0}$. These
massive vectors correspond to the broken non-compact generators of
the full gauge group.

\section{Moduli space of supersymmetric $\AdS_6$ vacua}\label{moduli}
In this section, we determine the flat directions of the scalar
potential $V$ which preserve all 16 supercharges.
These are the  moduli of the $\AdS_6$ backgrounds corresponding to supersymmetric marginal deformations of the five-dimensional superconformal field theories dual to the $\AdS_6$ vacua identified
in the previous section.

 Similar to the analysis of
\cite{AdS4_N4_Jan,AdS_7_N2_Jan,AdS5_N4_Jan},  a necessary condition for the existence of these moduli can be
determined by considering possible deformations of the supersymmetry
conditions \eqref{dPsi} near the $\AdS_6$ vacua. By
varying the conditions in \eqref{AdS_conditions1}, we find
\begin{eqnarray}
\delta (e^{4\sigma}A)=4\langle A\rangle\delta\sigma+e^{4\langle\sigma\rangle}\delta A&=&0,\label{moduli_condition1}\\
\delta C^I_{\phantom{I}t}=\delta D^I_{\phantom{I}t}=\delta B_t&=&0\,
.\label{moduli_condition2}
\end{eqnarray}
 We now introduce a parametization of the variation of the
coset representative with respect to the $4n$ scalars $\phi^{\alpha
I}$
\begin{equation}
\delta L^\Lambda_{\phantom{\Lambda}\alpha}=\langle
L^\Lambda_{\phantom{\Lambda}I}\rangle\delta \phi^{\alpha I},\qquad
\delta L^\Lambda_{\phantom{\Lambda}I}=\langle
L^\Lambda_{\phantom{\Lambda}\alpha}\rangle\delta \phi^{\alpha I}
\end{equation}
and their inverse
\begin{equation}
\delta (L^{-1})^\Lambda_{\phantom{\Lambda}\alpha}=-\langle
(L^{-1})^\Lambda_{\phantom{\Lambda}I}\rangle\delta \phi^{\alpha
I},\qquad \delta (L^{-1})^\Lambda_{\phantom{\Lambda}I}=-\langle
(L^{-1})^\Lambda_{\phantom{\Lambda}\alpha}\rangle\delta \phi^{\alpha
I}\, .
\end{equation}
Using these relations and the decomposition of indices
$\alpha=(0,r)$, we find
\begin{equation}
\delta L^\Lambda_{\phantom{\Lambda}0}=\langle
L^\Lambda_{\phantom{\Lambda}I}\rangle\delta \phi^{0I},\qquad
\delta L^\Lambda_{\phantom{\Lambda}i}=\langle L^\Lambda_{\phantom{\Lambda}I}\rangle\delta \phi^{iI},\qquad
\delta L^\Lambda_{\phantom{\Lambda}I}=\langle
L^\Lambda_{\phantom{\Lambda}0}\rangle\delta \phi^{0I}+ \langle
L^\Lambda_{\phantom{\Lambda}r}\rangle \delta\phi^{rI}
\end{equation}
and
\begin{equation}\begin{aligned}
\delta (L^{-1})^\Lambda_{\phantom{\Lambda}0}&=-\langle
(L^{-1})^\Lambda_{\phantom{\Lambda}I}\rangle\delta \phi^{0I},\qquad
\delta (L^{-1})^\Lambda_{\phantom{\Lambda}i}=-\langle (L^{-1})^\Lambda_{\phantom{\Lambda}I}\rangle\delta \phi^{iI},\\
\delta (L^{-1})^\Lambda_{\phantom{\Lambda}I}&=-\langle
(L^{-1})^\Lambda_{\phantom{\Lambda}0}\rangle\delta \phi^{0I}-
\langle (L^{-1})^\Lambda_{\phantom{\Lambda}r}\rangle
\delta\phi^{rI}\, .
\end{aligned}\end{equation}
 With the help of these relations, we can rewrite the conditions
\eqref{moduli_condition1} and \eqref{moduli_condition2}
as
\begin{eqnarray}
0&=&\delta(e^{4\sigma}A)=4e^{4\langle\sigma\rangle}\langle A\rangle\delta \sigma
+3e^{4\langle\sigma\rangle}\langle C_{Ir}\rangle \delta\phi^{rI},\\
0&=&\delta B_t=\langle C_{It}\rangle \delta \phi^{0I}+2\epsilon_{rtk}\langle D_{Ik}\rangle \delta\phi^{rI},\\
0&=&\delta
C^I_{\phantom{I}t}=2\epsilon^{trs}K_{[rIJ}\delta\phi_{s]J}-\epsilon^{trs}K_{r0s}\delta\phi^{0I}
-\epsilon^{trs}K_{ris}\delta\phi^{iI}
,\\
0&=&\delta
D^I_{\phantom{I}t}=K_{0It}\delta\phi^{0I}+K_{0IJ}\delta\phi^{tJ}-K_{0rt}\delta\phi^{rI}
\end{eqnarray}
where
\begin{equation}
K_{0IJ}=f_{\Lambda
\Sigma\Pi}L^\Lambda_{\phantom{r}0}(L^{-1})_I^{\phantom{s}\Sigma}L^\Pi_{\phantom{s}J},\qquad
K_{rIJ}=f_{\Lambda
\Sigma\Pi}L^\Lambda_{\phantom{r}r}(L^{-1})_I^{\phantom{s}\Sigma}L^\Pi_{\phantom{s}J}\,
.
\end{equation}
 Using the $\AdS_6$ conditions
\eqref{AdS_conditions1}, \eqref{D_C_conditions} and \eqref{B_condition} obtained in the
previous section, we find
\begin{equation}\label{modulieq}
\delta\sigma =0,\quad
K_{0\tilde{I}\tilde{J}}\delta\phi_{t\tilde{J}}=0,\quad
K_{rst}\delta\phi_{t\tilde{I}}=0,\quad
2\epsilon^{rst}K_{[r\hat{I}\hat{J}}\delta\phi_{s]\hat{J}}+K_{rst}\delta \phi_{t\hat{I}}=0\, .
\end{equation}
From these conditions, we immediately obtain
$\delta\phi_{t\tilde{I}}=0$ for $K_{rst}\neq 0$.

 The last equation in \eqref{modulieq} is similar to the one considered in
\cite{AdS4_N4_Jan,AdS_7_N2_Jan}, and it has been shown in
\cite{AdS4_N4_Jan} that this equation has general solutions of the
form
\begin{equation}
\delta\phi_{s\hat{I}}=K_{s\hat{I}\hat{J}}\lambda^{\hat{J}}\, .
\end{equation}
The remaining scalars that are not fixed by the above conditions are
$\delta\phi_{0\tilde{I}}$. We can readily recognize that
$\delta\phi_{s\hat{I}}$ and $\delta\phi_{0\tilde{I}}$ correspond to
Goldstone bosons of the symmetry breaking $G'\times G''\rightarrow
SO(3)\times H'\times H''$, with $H'\subset SO(m)$ and $H''\subset
SO(n-m)$ in the $\AdS_6$ vacuum. These massless scalars are eaten by
the massive gauge fields mentioned in the previous section. Thus, all of
the flat directions correspond to Goldstone bosons and no moduli exist.
This in turn is consistent with the fact  that there are no marginal deformations
preserving all supersymmetry in the dual five-dimensional SCFTs.
\section{Conclusions}\label{conclusion}
In this paper, we have analyzed the general conditions for the
existence of maximally supersymmetric $\AdS_6$ vacua in the $N=(1,1)$
half-maximal gauged supergravities in six dimensions. We have found
that three of the graviphotons have to gauge an $SU(2)_R$
R-symmetry while the forth one can be used to gauge a commuting
non-compact group. The fact that the $SU(2)_R$ R-symmetry must be
gauged is similar to the results in $d=4,6,7$.
This is in general a necessary condition for the existence of AdS
vacua as shown in \cite{Max_super_bg_Jan}. It is also consistent
with the important role played by the corresponding R-symmetry in
the dual field theories \cite{Seiberg_5Dfield}. Furthermore, all
vacua we have identified have no flat directions which preserve all supercharges
 corresponding to the absence of supersymmetric exactly
marginal deformations in the dual five-dimensional SCFTs.

 We end the paper by briefly giving some comments on the
$\mathbb{R}^+\times SO(4,n)$ covariant formulation in term of the
embedding tensor. As shown in \cite{Eric_Kac_Moody}, there are two
components of the embedding tensor given by $\xi^\Lambda$ and
$f_{\Lambda\Sigma\Gamma}$ as well as a massive deformation of the two-form field. The $\xi^\Lambda$ is involved in gauging of the $\mathbb{R}^+$ factor. Due to many similarities between the
six-dimensional $N=(1,1)$ gauged supergravity considered here and
the $N=2$ gauged supergravity in seven dimensions, we expect that the $\mathbb{R}^+$ gauging and the massive deformation could not be turned on simultaneously. Therefore, the existence of maximally supersymmetric $\AdS_6$ vacua would require $\xi^\Lambda=0$ as shown in \cite{AdS_7_N2_Jan} for the seven-dimensional case. It would be
of particular interest to explore this issue in particular to construct the complete gauging of $N=(1,1)$ supergravity in the embedding tensor formulation.
\newpage

\section*{Acknowledgement}

This work is supported in part by the German Science Foundation
(DFG) under the Collaborative Research Center (SFB) 676 ``Particles,
Strings and the Early Universe''. We would like to thank Nathan
Seiberg for useful correspondence. P. K. is grateful to the SFB,
Hamburg University for hospitality while most of this work has been
done. He is also supported by The Thailand Research Fund (TRF) under
grant RSA5980037.


\end{document}